\documentstyle[preprint,aps]{revtex}

\begin{document}

\title{Interface modes of two-dimensional composite structures}
\author{J. M. Pitarke$^1$\cite{byline}, F. J.
Garc\'\i a-Vidal$^2$, and J. B. Pendry$^3$} 
\address{$^1$Materia Kondentsatuaren Fisika Saila, Zientzi
Fakultatea, Euskal Herriko  Unibertsitatea,\\ 644 Posta kutxatila, 48080 Bilbo,
Basque Country, Spain\\
$^2$Departamento de F\'\i sica Te\'orica de la Materia Condensada,
Facultad de Ciencias,\\ Universidad Aut\'onoma de Madrid, 28049 Madrid, Spain\\
$^3$Condensed Matter Theory Group, The Blacket Laboratory, Imperial College,\\ 
London SW7 2BZ, United Kingdom}
\date\today
\draft
\maketitle

\begin{abstract}

The surface modes of a composite consisting of aligned metallic wires with
square cross sections are investigated, on the basis of photonic band structure
calculations. The effective long-wavelength dielectric response function is
computed, as a function of the filling fraction. The dependence of the optical
absorption on the shape of the wires and the polarization of light is
discussed, and the effect of sharp corners analyzed. The effect of the
interaction between the wires on the localization of surface plasmons is also
addressed.

\end{abstract}

\bigskip

\pacs{{\it Keywords}: Interface states; Light scattering; Plasmons;
Electron-solid interactions, scattering, diffraction.}




The interface electromagnetic modes of composite structures\cite{Bergman1} are
well-known to be of basic importance in the understanding of a
number of physical problems such as optical absorption\cite{Landauer}, van der
Walls attraction\cite{Langbein1}, surface-enhanced Raman
scattering\cite{SERS,FJ0}, and the response to incident charged
particles\cite{Ritchie1,Batson,Pitarke0}. Analytical descriptions of
normal electric modes have been limited to slabs and isolated
spheres\cite{Ruppin,Echenique} and spheroids\cite{Bohren}, while
numerical methods have been required for most shapes\cite{Knipp1}
including such simple cases as single cubes\cite{Fuchs,Langbein2} and
rectangles\cite{Langbein2}.

Dipolar modes in a system of identical spheres embedded in an otherwise
homogeneous host have been approximately described, for many years, from the
knowledge of the effective transverse dielectric function first derived by
Maxwell-Garnett\cite{MG} (MG) within a mean-field approximation valid for small
values of the volume occupied by the spheres. Also, many attempts have been made
to account, at large filling fractions, for higher multipole interactions, which
are absent in the MG theory\cite{Wood,Mc,Lamb,Tao,Bergman2}. Recently, new
techniques have been developed for solving Maxwell's equations in structured
materials\cite{Leung,Ho,Pendry1}. Based on these numerical solutions of Maxwell's
equations, exact calculations of the long-wavelength limit of the effective
transverse dielectric function of periodic dielectric\cite{Datta} and
metallic\cite{Moreno,FJ,Pitarke1} structures have been performed. From these
calculations, strengths and positions of surface modes of a composite made up of
long metallic cylinders have been obtained\cite{Pitarke1}, and the MG theory has
been shown to offer a good approximation as long as the distance between the
axis of neighboring cylinders is larger than twice the diameter of the cylinders.

In this paper we extend the work presented in Ref.\onlinecite{Pitarke1} to the
case of aligned metallic wires with square cross sections. We consider a
binary metal-dielectric mixture with a volume fraction $1-f$ of the insulator,
with a real positive dielectric constant $\epsilon_0$, and $f$ of a periodic
system of long metallic wires with square cross sections
arranged in a square array with lattice constant $a=xd$, as shown in Fig. 1.
The cross section of the wires is taken to be  small in comparison to the
wavelength of the electromagnetic excitation and large enough that a
macroscopic dielectric function $\epsilon(\omega)$ is ascribable to the wires.
Thus, the electromagnetic properties of the composite will be defined by a
single effective dielectric function,
\begin{equation}
\epsilon_{\rm eff}(\omega)={k^2c^2\over\omega^2},
\end{equation}
the wave vector $k$ being that corresponding to a Bloch wave propagating
through the composite. For simplicity, the magnetic permeabilities are assumed to
be equal to unity in both media.

The dielectric function of metals is nearly real and negative below the plasma
frequency down to frequencies comparable to the inverse of the conductivity mean
free time. Hence, negative values of $\epsilon(\omega)$ give rise
to a variety of surface resonances, which allow to express the
long-wavelength limit of the effective dielectric function of the composite in
the following form\cite{Bergman3,Milton}:
\begin{equation}     
\epsilon_{\rm eff}(\omega)=\epsilon_0\left[1-f\sum_\nu{B_\nu\over
u(\omega)-m_\nu}\right],
\end{equation}
where
\begin{equation}
u(\omega)=\left[1-\epsilon(\omega)/\epsilon_0\right]^{-1},
\end{equation}
$m_\nu$ ($0\le m_\nu<1$) is the depolarization factor associated with the
$\nu$th normal mode, and $B_\nu$ ($B_\nu\ge 0$) represent the corresponding
strengths, which all add up to unity:
\begin{equation}
\sum_\nu B_\nu=1.
\end{equation}
Similarly\cite{note0},
\begin{equation}     
\epsilon^{-1}_{\rm eff}(\omega)=\epsilon_0^{-1}\left[1+f\sum_\nu{C_\nu\over
u(\omega)-n_\nu}
\right],
\end{equation}
where $0<n_\nu\le 1$, $C_\nu\ge 0$, and
\begin{equation}
\sum_\nu C_\nu=1.
\end{equation}
Both depolarization factors and strengths in Eqs. (2) and (5) depend only on the
microgeometry of the composite material and not on the dielectric function of
the components.

Optical absorption is directly given by the imaginary part of the effective
dielectric function of Eq. (2), and coupling to charged particles is described
by the imaginary part of the effective inverse dielectric function\cite{note1}
of Eq. (5). In a homogeneous metal ($f=1$), there is only one mode strength
different from zero, with depolarization factors $m_0=0$ and $n_0=1$, where ${\rm
Re}\,\epsilon\to\pm\infty$ and
${\rm Re}\,\epsilon\approx 0$, respectively.
Thus, for a metal described by the Drude dielectric function,
$\epsilon(\omega)=1-\omega_p^2/\omega^2$, the bulk resonance at the plasma
frequency, $\omega_p$, can only be excited by penetrating charged particles. 

In the case of two-dimensional composite structures (see Fig. 1), homogeneous
along the axis of the inclusions, there are two different values
of $\epsilon_{\rm eff}(\omega)$ corresponding to different polarizations. If
the electromagnetic (e.m.) wave incident on the structure is polarized along the
axis of the inclusions ($s$-polarization), the presence of the interfaces does
not modify the electric field and the response of the composite is easily found
to be equivalent to that of a homogeneous medium having the
effective dielectric function of Eqs. (2) and (5) with only one mode strength
different from zero, $B_0=C_0=1$, and depolarization factors $m_0=0$ and
$n_0=f$\cite{note2,note3}. Thus, for this polarization the energy-loss function,
${\rm Im}[-\epsilon_{\rm eff}^{-1}(\omega)]$, shows a single peak at a reduced
plasma frequency. For a metal described by the Drude dielectric
function, the reduced plasma frequency is $\omega=\sqrt{f}\omega_p$.

Now we focus on the response to e.m. waves polarized within the plane normal to
the axis of the inclusions ($p$ polarization). For this polarization all
depolarization factors
$m_\nu$ and $n_\nu$ of Eqs. (2) and (5) are found to satisfy the
relation\cite{Pitarke1}
\begin{equation}
n_\nu=1-(D-1)m_\nu,
\end{equation}
where $D$ represents the dimensionality of the inclusions, i.e., $D=2$. As
long as only dipole surface modes can be excited, $B_1=C_1=1$, a combination of
Eqs. (2) and (5) with Eq. (7) results, for $D=2$, in the depolarization
factors $m_1$ and $n_1$ to be given by
\begin{equation}
m_1={1\over 2}(1-f)
\end{equation}
and
\begin{equation}
n_1={1\over 2}(1+f).
\end{equation}
Eqs. (2) and (5) with $B_1=C_1=1$ and the depolarization factors of Eqs. (8) and
(9) reduce to the two-dimensional MG effective dielectric
function\cite{note4}, as discussed in Ref.\onlinecite{Pitarke1}. We note that
multipolar modes can only be neglected in a few simple cases, in which the long
wave-length effective dielectric function is accurately given by the MG
approximation. These are the cases when the two-dimensional structure is made
of (a) a sparse ($f\to 0$) distribution of circular inclusions, and (b) a
dense ($f\to 1$) distribution of square inclusions.

For a single inclusion embedded in a host material with
dielectric constant $\epsilon_0$, both ${\rm Im}[\epsilon_{\rm eff}(\omega)]$
and ${\rm Im}[-\epsilon_{\rm eff}^{-1}(\omega)]$ are proportional to the
particle susceptibility, and satisfy the relation
\begin{equation}
{\rm Im}[\epsilon_{\rm eff}^{-1}(\omega)]={\rm Im}[\epsilon_{\rm
eff}(\omega)]/\epsilon_0^2,
\end{equation}
i.e., the spectral representations of Eqs. (2) and (5) coincide. Thus, one
finds from Eq. (7) that, in the case of single inclusions, for any given mode
with
$m_\nu<1/D$ there exists another mode with depolarization factor
$m_\nu'>1/D$ such that
\begin{equation}
m_\nu'=1-(D-1)m_\nu.
\end{equation}
In particular, for single circular inclusions only dipole surface modes can
be excited, depolarization factors being $m_1=n_1=1/2$, and
both the optical absorption and the electron energy loss exhibit a single
strong maximum at the dipole resonance where
$u=1/2$, i. e., $\epsilon(\omega)+\epsilon_0=0$. For single inclusions with
shapes other than circular, this resonance splits into several peaks that occur
at a set of frequencies according to Eq. (11).

In order to give a full quantitative description of all surface modes in our
two-dimensional composite structure, we have followed the method developed in
Ref.\onlinecite{Pendry1} for calculating dispersion relationships of Bloch waves
in structured materials, and we have computed the effective dielectric function
from Eq. (1). The insulating component of the composite has been chosen to have
a real dielectric constant of $\epsilon_0=1$ and the dielectric properties of
the metal inside the inclusions have been modeled by a Drude dielectric
function of the form
\begin{equation}
\epsilon(\omega)=1-{\omega_p^2\over\omega(\omega+{\rm i}\gamma)},
\end{equation}
where $\omega_p$ represents the bulk plasma frequency of the metal and $\gamma$
is an inverse electron relaxation time. The plasma frequency of the
conduction electrons in aluminum has been used,
$\hbar\omega_p=15.8{\rm eV}$, and the parameter $\gamma$ has been chosen to
be small enough so as to distinguish all multipolar
excitations: $\gamma=0.1{\rm eV}$.

Figs. 2 and 3 show our numerical calculations of ${\rm Im}[\epsilon_{\rm
eff}(\omega)]/f$ and ${\rm Im}[-\epsilon_{\rm eff}^{-1}(\omega)]/f$, as a
function of the frequency $\omega$, for various values of the ratio $x$
between the lattice constant and the side of the wires. For $x\ge 6$ our
results are almost insensitive to the precise value of $x$, and they nearly
coincide with the effective dielectric function of single wires. In this
(dilute) limit, ${\rm Im}[\epsilon_{\rm eff}(\omega)]$ and ${\rm
Im}[-\epsilon_{\rm eff}^{-1}(\omega)]$ coincide, and they both exhibit several
resonances satisfying Eq. (11) (see the insets of Figs. 2 and 3)\cite{note5}. 

At higher concentrations of metal, both the optical absorption (${\rm
Im}[\epsilon_{\rm eff}(\omega)]$) and the electron energy loss (${\rm
Im}[-\epsilon_{\rm eff}^{-1}(\omega)]$) show a dominant dipole-like mode
and a band of multipole resonances. We note (see Figs. 2 and 3) that as $x$
decreases the dipole-like resonance in ${\rm Im}[\epsilon_{\rm eff}(\omega)]/f$
and ${\rm Im}[-\epsilon_{\rm eff}^{-1}(\omega)]/f$ is shifted towards lower and
higher frequencies, respectively, as a consequence of the interaction between
one surface and another. This result, already predicted within the MG
approximation (see Eqs. (8) and (9)), is also found in the case of circular
inclusions\cite{Pitarke1}. Also, it is interesting to notice that as long as $x$
decreases multipole resonances tend to be negligible, thus
dipolar depolarization factors approaching those predicted by Eqs. (8) and (9).

In the case of circular wires that are touching ($f=0.785$) the metal forms a
connected medium, and the so-called energy-loss function 
${\rm Im}[-\epsilon_{\rm eff}^{-1}(\omega)]$ shows, therefore, a bulk plasmon
excitation at $\omega=\omega_p$ ($n_1=1$). Besides this peak, there is also a
band in ${\rm Im}[-\epsilon_{\rm eff}^{-1}(\omega)]$ from multipole
contributions to the effective response at $\omega_\nu=\sqrt{n_\nu(f)}\omega_p$
($\nu=2,...;n_\nu<1$), which prevents our computed effective dielectric
function to coincide with the MG approximation (see Ref.\onlinecite{Pitarke1}).
However, the presence of flat surfaces with sharp corners, as in the case of
square inclusions, results in all multipolar strengths to be negligible for
$f>0.75$ ($x<1.15$); thus, our computed effective dielectric function is
well described, in this case and for $x<1.15$, by the MG approximation.

Finally, we have represented in Fig. 4 universal curves (solid lines) for the
dipolar mode depolarization factors and strengths, versus the filling fraction,
as determined with use of Eqs. (13) and (14) from our computed effective
dielectric function of a system of infinitely long wires with square cross
sections. An inspection of this figure indicates that the trend with increasing
filling fraction is for the dipolar peaks in ${\rm Im}[\epsilon_{\rm
eff}(\omega)]$/${\rm Im}[-\epsilon_{\rm eff}^{-1}(\omega)$ to approach the MG
results of Eqs. (8) and (9), also plotted in this figure (dotted lines);
furthermore, the MG results nearly coincide with our numerical results for
$f>0.8$ ($x<1.1$). Associated with the deviation, for smaller filling
fractions, of the actual dipolar mode positions from the MG results is the
reduction in the dipolar mode strengths $B_1$ and $C_1$, which are represented
in the inset of Fig. 4. These mode strengths have
been determined by
\begin{equation}
B_\nu={\sqrt{m_\nu}\over f\,H}{\rm Im}[\epsilon_{\rm eff}(\omega_\nu)]
\end{equation}
and
\begin{equation}
C_\nu={\sqrt{n_\nu}\over f\,H}{\rm Im}[-\epsilon_{\rm eff}^{-1}(\omega_\nu)],
\end{equation}
where $\omega_\nu$ is the frequency associated with the $\nu$th
normal mode, and $H$ represents the peak height in the bulk energy-loss
function [in the case of the Drude dielectric function of Eq. (12),
$m_\nu,n_\nu=\omega_\nu^2/\omega_p^2$ and $H=\omega_p/\gamma$].

In summary, we have presented, on the basis of {\it ab initio} numerical
solutions of Maxwell's equations in structured materials, exact
calculations of the effective long-wavelength dielectric response function of
a periodic system of long  metallic wires with square cross
sections. We have investigated the dependence of both the optical absorption and
the electron energy loss on the shape of the wires and the polarization of
light. In the case of $s$ polarization the effective dielectric function does
not depend on the shape of the wires and there is only one mode strength
different from zero. In the case of $p$ polarization it had already been
concluded\cite{Pitarke1} that MG results are good as long as the distance
between the axis of neighboring circular cylinders is {\it larger} than twice
the diameter of the cylinders. Now we have concluded that in the case of
two-dimensional periodic structures consisting of square inclusions MG results
are reliable as long as the ratio between the lattice constant and the side of
the inclusions is {\it smaller} than $x\sim 1.1$. Both the presence of sharp
corners in single wires and the interaction between circular wires result in
absorption and energy-loss peaks to be broadened by the existence of multipolar
surface resonances.

J. M. P. acknowledges partial support by the Basque Unibertsitate eta Ikerketa
Saila and the Spanish Ministerio de Educaci\'on y Cultura.

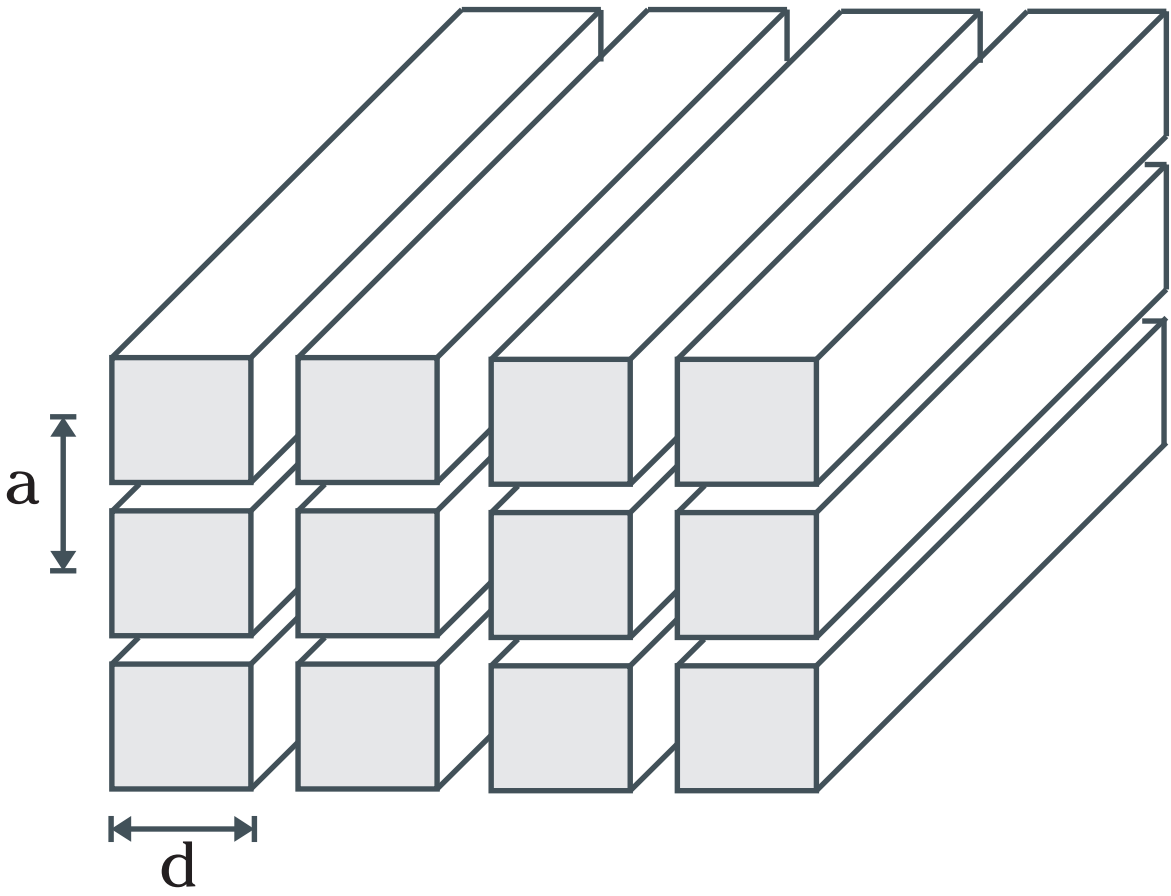
\begin{figure}
\caption{Periodic system of metallic wires with square cross sections of side
$d$, arranged in a square array with lattice constant $a$. The wires are
infinitely long in the $y$ direction.}
\end{figure}

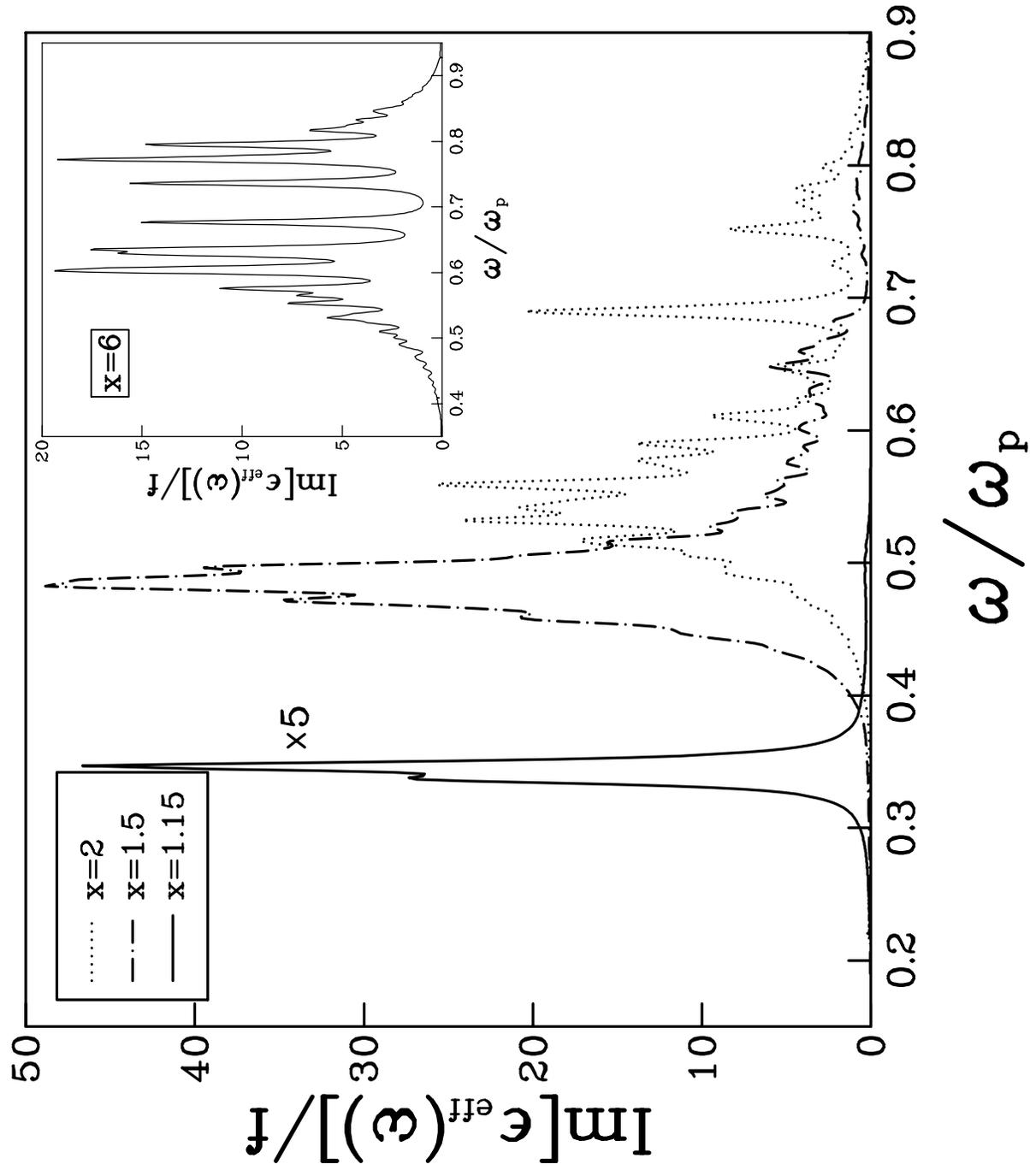
\begin{figure}
\caption{Imaginary part of the effective long-wavelength dielectric function of
the periodic system described in Fig. 1, for $p$ polarized electromagnetic
excitations and various values of the volume filling fraction: $25.0\%$
(dotted line), $44.4\%$ (dashed line), and $75.6\%$ (solid line). The results
obtained for a volume filling fraction of $2.3\%$ are represented in the
inset.}
\end{figure}

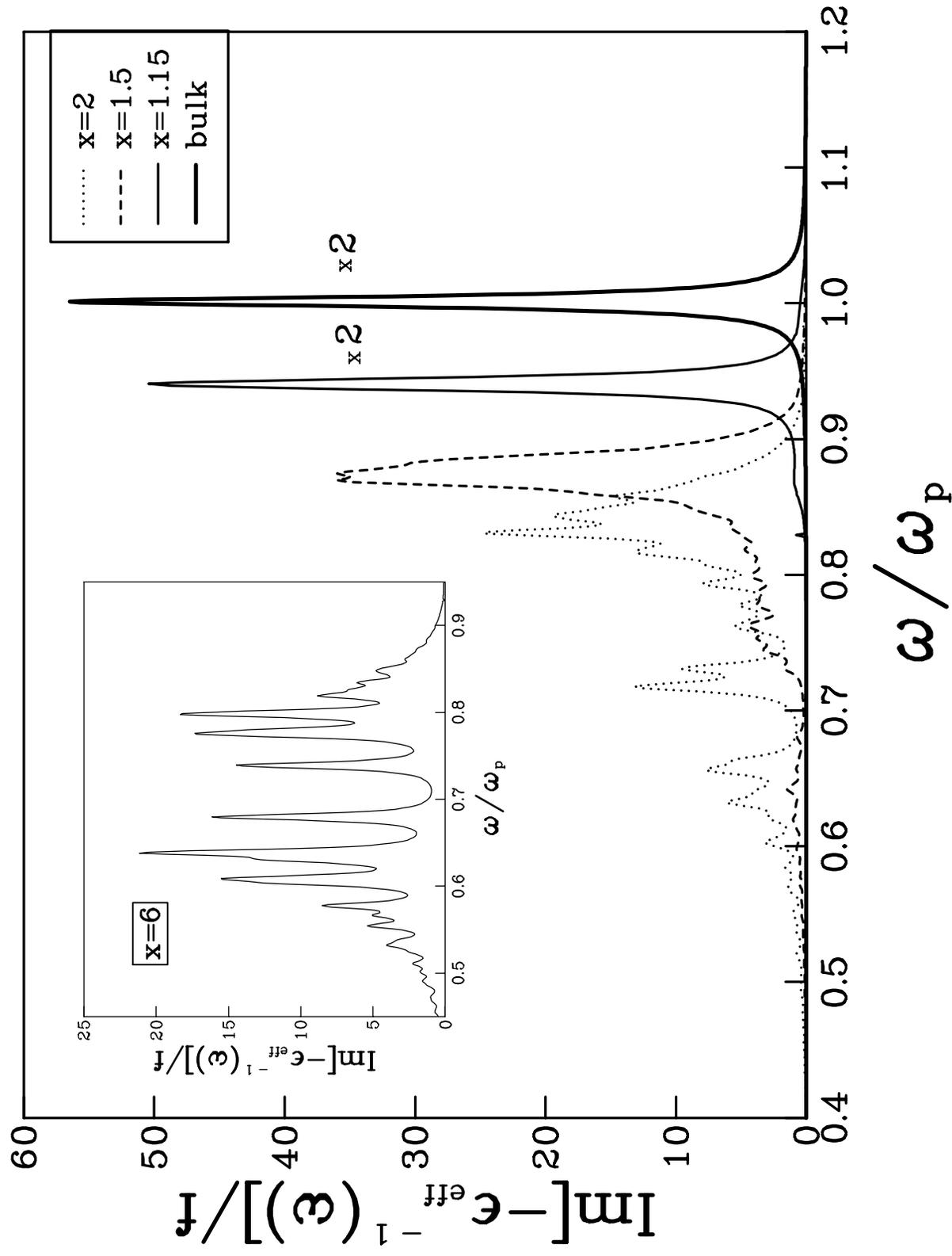
\begin{figure}
\caption{Same as in Fig. 2, for the effective energy-loss function. The bulk
energy-loss function ($x=1$) is represented by a thick solid line.}
\end{figure}

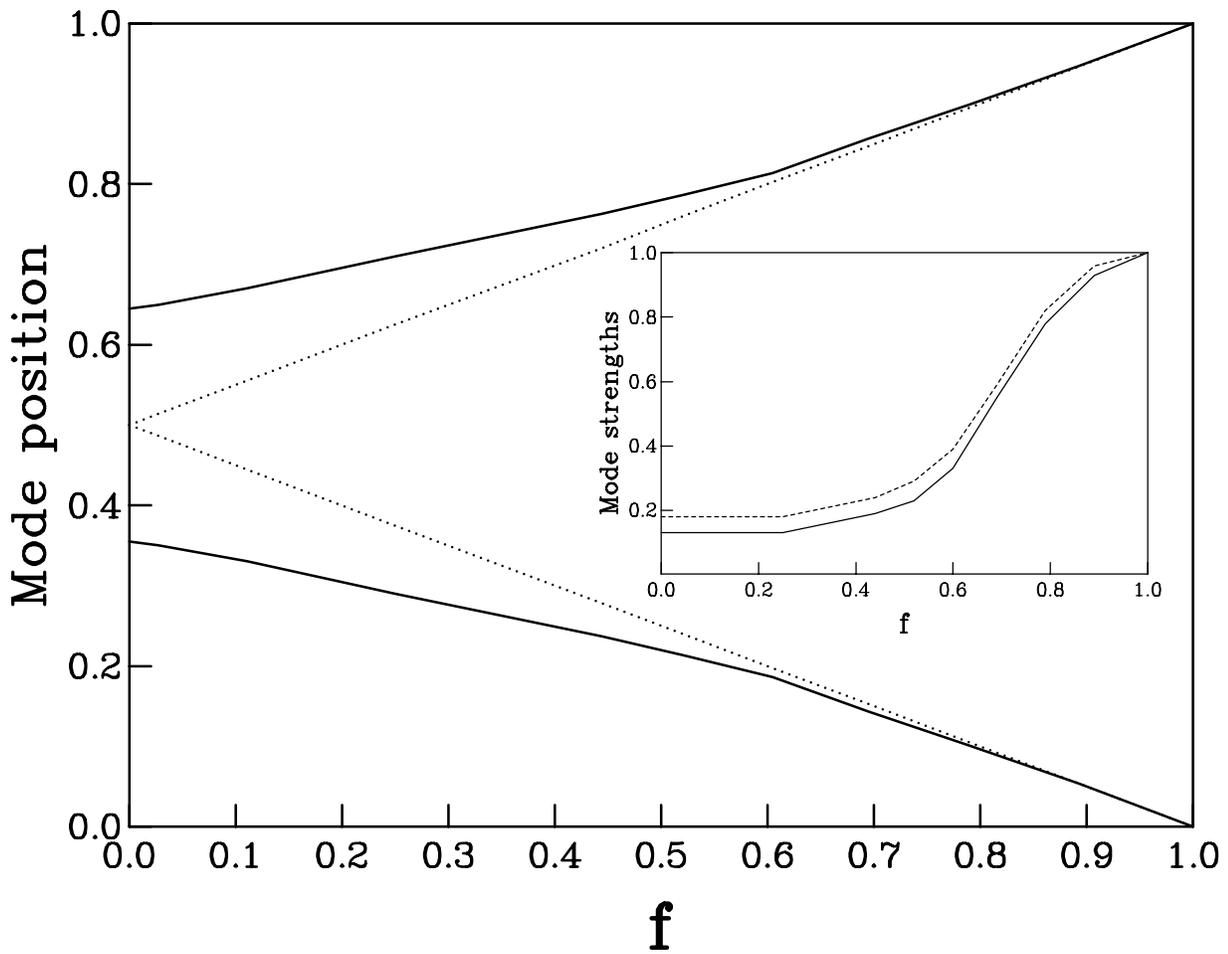
\begin{figure}
\caption{Dipolar mode positions (depolarization factors), as a function of
the volume filling fraction, for wires with square cross sections. Solid
lines: our numerical results for $m_1$ (the line below $0.5$) and $n_1$
(the line above $0.5$). Dotted lines represent the MG depolarization
factors given by Eqs. (8) and (9). Dipolar mode strengths are represented
in the inset: Solid and dashed lines represent the coefficients $B_1$ and
$C_1$ entering Eqs. (2) and (5), respectively. The MG dipolar mode strengths,
$B_1$ and $C_1$, are both equal to unity.}
\end{figure} 


\begin{references}
\bibitem{note0} The minus sign in front of $f$ in Eq. (4) of
Ref.\onlinecite{Pitarke1} must be corrected and replaced by a plus sign, as in
the present paper.
\bibitem{note2} The same result is found in the case of plane parallel slabs
aligned along the electric field (see, e.g., Ref.\onlinecite{Bergman1}). If
the electric field is perpendicular to the slabs, there is also only one mode
strength different from zero, but now with depolarization factors $m_0=1-f$ and
$n_0=1$.
\bibitem{note4} For three-dimensional composite structures, $D=3$, a
combination of Eqs. (2) and (5) with Eq. (7) results, if all multipolar modes
are neglected, in the dielectric function originally derived by
Maxwell-Garnett\cite{MG} for a system of spherical particles.
\bibitem{note5} Although the results presented in Figs. 2 and 3 are all
well converged for $x\le 2$, the results presented for $x=6$ (corresponding
to a sampling mesh of $120\times 1\times 120$) might slightly
differ at $\omega\sim\omega_p/\sqrt{2}$ from well-converged results.
\end{references}
\end{document}